\documentclass[letterpaper,10pt,conference]{ieeeconf}

\IEEEoverridecommandlockouts
\usepackage{amsmath} 
\usepackage{amssymb}  
\usepackage[utf8]{inputenc}
\usepackage{color}
\usepackage{tikz}
\usetikzlibrary{shapes,arrows.meta}
\usepackage{xspace}
\usepackage{siunitx}
\usepackage{cite}
\usepackage{bm,fixmath}
\usepackage[T1]{fontenc}

\usepackage{pythonhighlight}

\usepackage{graphicx}
\graphicspath{{./image/}}

\tikzset{agent/.style={circle, draw, inner sep=8pt}}

\usepackage{algorithm}
\usepackage{algpseudocode}

\algrenewcommand\algorithmicrequire{\textbf{Input:}}

\DeclareMathOperator*{\argmin}{arg\,min}
\DeclareMathOperator{\prox}{prox}

\DeclareMathOperator{\proj}{proj}

\newtheorem{remark}{Remark}

\newtheorem{example}{Example}

\newcommand{\norm}[1]{\left\lVert#1\right\rVert}

\newcommand{\N}{\mathbb{N}}
\newcommand{\R}{\mathbb{R}}

\newcommand{\bv}{\mathbold{b}}
\newcommand{\cv}{\mathbold{c}}
\newcommand{\e}{\mathbold{e}}

\newcommand{\x}{\mathbold{x}}
\newcommand{\y}{\mathbold{y}}

\newcommand{\Am}{\mathbold{A}}
\newcommand{\Bm}{\mathbold{B}}

\newcommand{\Ts}{T_\mathrm{s}}
\newcommand{\Nc}{N_\mathrm{C}}
\newcommand{\Np}{N_\mathrm{P}}

\renewcommand{\baselinestretch}{0.96}

\newcommand{\tvopt}{\texttt{tvopt}\xspace}
\newcommand{\code}[1]{\pyth{#1}\xspace}

\title{\LARGE \bf
\tvopt: A Python Framework for Time-Varying Optimization
}
\author{Nicola Bastianello
\thanks{N. Bastianello is with the Department of Information Engineering (DEI), University of Padova, Italy. {\tt\scriptsize bastian4@dei.unipd.it}.}
}

\begin{document}

\maketitle

\begin{abstract}
This paper introduces \tvopt, a Python framework for prototyping and benchmarking time-varying (or online) optimization algorithms. The paper first describes the theoretical approach that informed the development of \tvopt. Then it discusses the different components of the framework and their use for modeling and solving time-varying optimization problems. In particular, \tvopt provides functionalities for defining both centralized and distributed online problems, and a collection of built-in algorithms to solve them, for example gradient-based methods, ADMM and other splitting methods. Moreover, the framework implements prediction strategies to improve the accuracy of the online solvers. The paper then proposes some numerical results on a benchmark problem and discusses their implementation using \tvopt. The code for \tvopt is available at \cite{code}.
\end{abstract}

\section{Introduction}\label{sec:introduction}
In recent years, \emph{time-varying (or online) optimization} has received increasing interest from the optimization, control, and learning communities \cite{shalev-shwartz_online_2011,dallanese_optimization_2020,simonetto_time-varying_2020}. Indeed, in many applications technological advances brought about a shift from traditional optimization problems to problems that have a dynamic nature, \emph{e.g.} because they depend on streaming sources of data. Static optimization techniques then need to be revisited and adapted in order to provided reliable, on-the-fly algorithms for solving time-varying problems.

The goal of this paper is to propose \tvopt, a framework written in Python for prototyping and benchmarking of online optimization algorithms, and to facilitate this shift from a static to a dynamic optimization context. The idea indeed is to provide all the necessary tools to model time-varying optimization problems, and to implement suitable solution algorithms and analyze their performance.

Formally, time-varying optimization problems can be modeled as follows:
\begin{equation}\label{eq:continuous-time-problem}
	\x^*(t) = \argmin_{\x \in \R^n} F(\x; t)
\end{equation}
where $F : \R^n \times R_+ \to \R \cup \{ +\infty \}$ is a cost function that varies over time. For example, we may have $F(\x;t) = \norm{\Am \x - \bv(t)}^2 / 2$ to solve a regression task that employs time-varying observations $\bv(t) = \Am \y(t) + \e(t)$ of a signal $\y(t)$, affected by additive noise $\e(t)$.

\noindent We are also interested in the \emph{distributed} counterpart of~\eqref{eq:continuous-time-problem}, defined as
\begin{equation}\label{eq:distributed-continuous-time-problem}
\begin{split}
	\x^*(t) &= \argmin_{x_i \in \R^n} \sum_{i = 1}^{N} F_i(x_i; t) \\
	&\text{s.t.} \ x_i = x_j,\ \text{if $i$, $j$ connected}
\end{split}
\end{equation}
where $N$ is the number of agents cooperating towards the solution of~\eqref{eq:distributed-continuous-time-problem}, and each $F_i$ is the private local cost of agent $i = 1, \ldots, N$. For example, a multi-agent system of robots may encode a coordination task (\emph{e.g.} moving in formation) as the distributed optimization problem, which is inherently time-varying due to the dynamic nature of the system.

Different online optimization algorithms have been proposed, both for centralized problems \emph{e.g.} \cite{hall_online_2015,simonetto_class_2016}, and in decentralized scenarios \cite{ling_decentralized_2014,simonetto_decentralized_2017,bastianello_distributed_2020_b}. An interesting approach to developing online algorithms is that of \emph{prediction-correction}, proposed in \cite{simonetto_class_2016,simonetto_prediction-correction_2017} and extended in \cite{bastianello_primal_2020}. The main idea is to exploit past information on the problem to improve the solution accuracy of the algorithm.

Applications in which online algorithms are required range from signal and image processing, to control, and smart grids.

\noindent We refer to the surveys \cite{dallanese_optimization_2020,simonetto_time-varying_2020} for an in-depth literature review and a discussion of the different applications.

\paragraph*{Contribution}
This paper describes the \tvopt framework and the theoretical approach that informed its design. The paper describes the main features of \tvopt, which can be summarized as follows:
\begin{itemize}
	\item \emph{problem modeling}: the framework provides an object-oriented approach to modeling and defining online optimization problems, both the costs and constraints;
	
	\item \emph{decentralized problems}: moreover, \tvopt offers tailored tools to model multi-agent networks and decentralized problems;
	
	\item \emph{solvers}: \tvopt implements widely used solution algorithms for different classes of unconstrained and constrained problems, both centralized and decentralized.
\end{itemize}

The paper also presents some results of numerical simulations performed on a benchmark problem, and shows and discusses their implementation using the tools of \tvopt.

\paragraph*{Paper organization}
Section~\ref{sec:p-c-framework} discusses the theoretical approach that informs the design of \tvopt. Section~\ref{sec:framework} describes the main components of the framework and their use for online optimization, while section~\ref{sec:distributed} presents additional tools for simulating distributed online problems. Section~\ref{sec:example} concludes with a numerical example implemented using \tvopt, and some simulations results.

\section{Time-Varying Optimization}\label{sec:p-c-framework}
In this section we review the approach to time-varying optimization that informed the design of the \tvopt framework. We refer the interested reader to the theoretical framework developed in \cite{bastianello_primal_2020} and the surveys \cite{dallanese_optimization_2020,simonetto_time-varying_2020} for more details.

\subsection{Problem formulation}
At the foundation of \tvopt is a \emph{discrete-time} approach to online optimization, see \emph{e.g.} \cite{dallanese_optimization_2020}, which samples~\eqref{eq:continuous-time-problem} in the following sequence of (static) problems:
\begin{equation}\label{eq:discrete-time-problem}
	\x^*(t_k) = \argmin_{\x \in \R^n} F(\x; t_k)
\end{equation}
where $t_k$, $k \in \N$, are the sampling instants and $\Ts = t_{k+1} - t_k$ is a chosen sampling time. This approach is opposed to a \emph{continuous-time} one, see \emph{e.g.} \cite{fazlyab_prediction-correction_2018}.

The goal then is to track the \emph{optimal trajectory} given by the sequence $\{ \x^*(t_k) \}_{k \in \N}$ of minimizers of the sampled costs. However, the dynamic nature of the problem implies that the optima can be tracked only approximately, since a limited computational time (upper bounded by $\Ts$) is available to solve each problem in the sequence.

In order to illustrate this framework, consider the following examples.

\begin{example}
Let $\y(t)$ be a signal to be reconstructed from the noisy observations $\bv(t_k) = \Am \y(t_k) + \e(t_k)$, with $\e(t_k)$ denoting random noise. In this case we can define $F(\x;t_k) = f(\x; t_k) + g(\x)$ where $f(\x; t_k) = (1/2) \norm{\Am \x - \bv(t_k)}^2$ fits the observed data, and $g(\x) = w \norm{\x}_1$ promotes sparsity.
\end{example}

\begin{example}
In model predictive control (MPC), a control law is designed by solving a sequence of optimization problems which vary over time, since they depend on the states of a dynamical system. Thus MPC can be cast as a time-varying optimization problem and solved using tools developed in this framework. See \cite{simonetto_time-varying_2020} for an overview.
\end{example}

\subsection{Solvers}
As remarked above, each problem in the sequence~\eqref{eq:discrete-time-problem} needs to be (approximately) solved. We thus introduce the concept of \emph{solver}, by which we mean any recursive algorithm that can be applied to the sampled problems.

Due to the limited computational time that is available between the observation of a problem and the next, in general we cannot solve exactly each problem in the sequence. Rather, we apply a finite number of steps of the solver to each sampled problem, denoted by $\Nc \in \N$. This yields an approximate solution of the problems and thus leads to an approximate tracking of the optimal trajectory.

\begin{example}
Depending on the structure of the problem, a wide array of solvers can be used. For example, if $F(\x; t_k)$ is smooth, then a gradient method is a suitable solver. Or, for a composite problem, $F(\x; t_k) = f(\x; t_k) + g(\x; t_k)$, we can apply a proximal gradient method.
\end{example}

\subsection{Prediction}
As proposed in \cite{simonetto_class_2016,simonetto_prediction-correction_2017} and further extended in \cite{bastianello_primal_2020}, the knowledge of past sampled problems can be exploited to improve the tracking of the optimal trajectory. Indeed, the information collected up to time $t_k$ can be used to shape a \emph{prediction} of the (as yet unobserved) problem at time $t_{k+1}$. Then an approximate solution of the predicted problem can be used to \emph{warm-start} the solver\footnote{That is, the approximate prediction solution is used as initial condition for the solver applied to problem at time $t_{k+1}$.} when applied to the actual problem. As proved in \cite{bastianello_primal_2020}, this approach allows to reduce the tracking error.

\begin{example}\label{ex:prediction}
A very simple prediction strategy is to choose $\hat{F}(\x; t_{k+1}) = F(\x; t_k)$, where $\hat{F}$ denotes the prediction. Another, called \emph{extrapolation}, chooses $\hat{F}(\x; t_{k+1}) = 2 F(\x; t_k) - F(\x; t_{k-1})$.
\end{example}

\vspace{0.3cm}

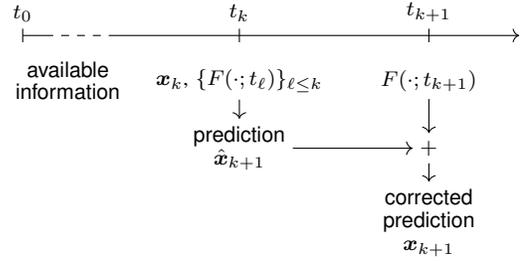
\begin{figure}[!ht]
\centering
\footnotesize
\begin{tikzpicture}[scale=0.6]

	\draw (-3,0) -- (-2.5,0);
	\draw[dashed] (-2.5,0) -- (-1,0);
	\draw[->] (-1,0) -- (8,0);
	\draw (-3,-0.15) -- (-3,0.15) node[above] {$t_0$};
	\draw (1.8,-0.15) -- (1.8,0.15) node[above] {$t_{k}$};
	\draw (6,-0.15) -- (6,0.15) node[above] {$t_{k+1}$};
	
	\node[align=center] at (-2,-1) {\textsf{available}\\\textsf{information}};
	\node (tk) at (1.8,-1) {$\x_k$, $\{ F(\cdot; t_\ell) \}_{\ell \leq k}$};
	\node (tk1) at (6,-1) {$F(\cdot; t_{k+1})$};
	
	\node[align=center] (pr) at (1.8,-2.5) {\textsf{prediction}\\$\hat{\x}_{k+1}$};
	\node[align=center] (co) at (6,-4.15) {\textsf{corrected}\\\textsf{prediction}\\$\x_{k+1}$};
	\node (plus) at (6,-2.5) {$+$};
	
	\path[->] (tk) edge (pr)
  (pr) edge (plus)
  (tk1) edge (plus)
  (plus) edge (co);
		
\end{tikzpicture}
\caption{The prediction-correction scheme.}
\label{fig:prediction-correction-scheme}
\end{figure}

To conclude this section, we detail the steps of a prediction-correction method for solving~\eqref{eq:discrete-time-problem}, see Figure~\ref{fig:prediction-correction-scheme}, and refer to \cite{bastianello_primal_2020} for further details.

\begin{enumerate}
	\item \emph{Initialization}: choose the sampling time $\Ts$, a prediction strategy, and the solver and its parameters.
\end{enumerate}
At each sampling time $t_k$, $k \in \N$:
\begin{enumerate}
	\setcounter{enumi}{1}
	
	\item \emph{Prediction}: \emph{(i)} predict the problem by computing $\hat{F}(\cdot; t_k)$, and \emph{(ii)} solve it approximately with $\Np \in \N$ steps of the solver, which yields the prediction $\hat{\x}_{k+1}$.
	
	\item \emph{Correction}: \emph{(i)} sample the new problem at time $t_{k+1}$, and \emph{(ii)} solve it approximately with $\Nc \in \N$ steps of the solver, using $\hat{\x}_{k+1}$ as initial condition. The result is denoted by $\x_{k+1}$.
\end{enumerate}

The framework delineated above can be particularized by omitting either the prediction or correction step. We remark that omitting the correction step yields the approach usually employed in \emph{online learning} \cite{shalev-shwartz_online_2011}.

\section{\tvopt Framework}\label{sec:framework}
In this section we describe the main components of \tvopt and their application to online optimization as reviewed in the previous section; see \cite{documentation} for the full documentation. The framework can be conceptually divided in the following:
\begin{itemize}
	\item \emph{Problem formulation}: the sub-modules \code{sets} and \code{costs}, which implement an object-oriented approach to defining time-varying problems. The sub-module \code{networks} can be used alongside the previous two to define distributed problems.
	
	\item \emph{Prediction}: the sub-module \code{prediction} is provided for approximating future problems based on the problems observed in the past.
	
	\item \emph{Solvers}: the sub-modules \code{solvers} and \code{distributed_solvers} implement a wide range of solvers that can be applied to~\eqref{eq:discrete-time-problem}.
\end{itemize}

In the following we describe in more details the sub-modules of \tvopt, while section~\ref{sec:distributed} will discuss the specific tools implemented for online distributed optimization.

\subsection{Sets}
This sub-module implements the \code{Set} objects which are used to define the domain of the cost objects. In particular, a \code{Set} is defined as a subset of $\R^{n_1 \times n_2 \times \ldots}$, for some $n_1, n_2, \ldots \in \N$. \code{Set} objects are then characterized by the dimensions $n_1, n_2, \ldots$ of the underlying space, which are stored in the attribute \code{shape}.

In this sub-module and in the following the unknown $\x$ of problem~\eqref{eq:discrete-time-problem} is modeled as a Numpy \code{ndarray} of proper size \cite{harris_array_2020}. In \tvopt, then, sets are built to be compatible with NumPy's \code{ndarrays}, and to use their broadcasting rules.

\begin{remark}
We remark that the most commonly used domains in online optimization are $\R^n$ and $\R^{n_1 \times n_2}$, the latter for example can be used for image processing without the need to vectorize. The definition of \code{Set} objects with more than two dimensions can however be useful as well, for example in the distributed scenario discussed in section~\ref{sec:distributed}.
\end{remark}

The other element of a \code{Set} definition is its \code{contains} method, which returns \code{True} if an input $\x$ is in the \code{Set}, \code{False} otherwise. When defining a \code{Set}, the \code{projection} method should also be implemented, which, given an input $\x$, returns its projection onto the set, defined as
$$
	\proj_{\mathbb{C}}(\x) = \argmin_{\y \in \mathbb{C}} \norm{\y - \x}^2
$$
with $\mathbb{C}$ the \code{Set}.

Finally, \code{Set} objects provide a \code{check_input} method which verifies if a given array \code{x} fits the dimension of the set (possibly reshaping it). This is useful to implement validity checks on the inputs inside a \code{Cost} method (see section~\ref{subsec:costs}).

\paragraph*{Operations}
The \code{contains} method can be accessed using the Python reserved keyword \code{in}. \code{Set} objects can be modified via the \code{scale} and \code{translate} methods. We can also define intersections of \code{Sets} by summing them (that is, using the \code{+} operator), in which case an approximate projection onto the intersection is implemented using the method of alternating projections (MAP) \cite{reich_projection_2015}\footnote{We remark that MAP returns a point in the intersection, not the actual projection. However, in practice MAP is faster than methods that are proven to return the projection, see \cite{reich_projection_2015}.}.

\paragraph*{Built-ins}
Different \code{Sets} are implemented, for example: the whole space $\R^{n_1 \times n_2 \times \ldots}$, ball and box sets, and half-spaces. A particular built-in set is \code{T}, which defines the set $\{ t_k \in \R_+, \ k \in \N \}$ of sampling instants.

\subsection{Costs}\label{subsec:costs}
The sub-module \code{costs} implements the \code{Cost} object to define time-varying cost functions
$$
	F : \R^{n_1 \times n_2 \times \ldots} \times \R_+ \to \R \cup \{ +\infty \}
$$
or, as a sub-case, static costs. A cost is characterized by the \code{dom} and (optionally) \code{time} attributes, which point to \code{Sets} for $\R^{n_1 \times n_2 \times \ldots}$ and the sampling times $\{ t_k \in \R_+, \ k \in \N \}$.

\code{Cost} objects are then defined by the \code{function}, \code{gradient} and \code{hessian} methods, where \code{gradient} returns a (sub-)gradient evaluation and \code{hessian} is implemented only if the cost is twice differentiable.

\noindent For example, we evaluate the (sub-)gradient of a function \code{F} as \code{F.gradient(x, t)}.

The costs then provide a \code{proximal} method, which computes:
$$
	\prox_{\rho F(\cdot;t_k)}(\x) = \argmin_{\y \in \R^{n_1 \times n_2 \times \ldots}} \{ F(\y;t_k) + \norm{\y - \x}^2 / (2 \rho) \}
$$
using either a gradient or Newton method, depending on the smoothness of $F$. If a closed form proximal is available, \emph{e.g.} for quadratic costs or $\ell_1$ norms, then this method should be overwritten.

Time-varying costs provide the \code{time_derivative} method which computes, using backward finite differences \cite{quarteroni_numerical_2007}, derivatives of $F$ (or of its gradient and Hessian) w.r.t. time. For example the time-derivative of the gradient is approximated by
$$
	\hat{\nabla}_{t\x} F(\x; t_k) = \left( \nabla_\x F(\x; t_k) - \nabla_\x F(\x; t_{k-1}) \right) / \Ts.
$$
Further, the costs have a \code{sample} method that returns a static cost representing $F$ at a chosen sampling instant $t_k$.

\paragraph*{Operations}
Costs can be scaled by a scalar and elevated to a given power, and they can be summed and multiplied by other costs.

\paragraph*{Built-ins}
Some of the \code{Cost} objects that are implemented are $\norm{\cdot}_1$, $\norm{\cdot}_{\infty}$, quadratic cost, Huber loss, and the indicator function of any given \code{Set}. The benchmark dynamic costs proposed in \cite[section~IV.A]{simonetto_class_2016} and \cite[section~III.B]{zhang_zeroing_2019} are implemented. Dynamic costs can also be defined from a sequence of static costs using \code{DiscreteDynamicCost}.

\subsection{Prediction}
The sub-module \code{predictions} implements the \code{Prediction} object for approximating future costs from previously sampled costs. The object stores a dynamic \code{Cost} to be predicted and, through the method \code{update}, uses information on the cost up to a specified time $t_k$ to shape a prediction. The object behaves like a static cost, in the sense that it exposes the \code{function}, \code{gradient}, \emph{etc.} methods of the (static) predicted cost.

\paragraph*{Built-ins}
The sub-module implements the Taylor expansion-based and extrapolation-based prediction strategies studied in \cite{bastianello_primal_2020}.

\subsection{Solvers}\label{subsec:solvers}
The sub-module \code{solvers} implements a selection of algorithms for solving different classes of static problems. The solvers are Python functions that are passed a static problem (in the form of a dictionary), the number of iterations to be applied, and any required parameters, such as step-sizes. The functions then return an approximate solution.

All the solvers provided by \tvopt are not tailored to any specific choice of cost functions. Instead, they are defined to exploit the common template of \code{Cost} objects by calling \emph{e.g.} their \code{gradient}, without needing to know how \code{gradient} is actually computed.

\begin{remark}
Notice that the modular design of solvers allows to define costs that for example inexactly compute \code{gradient} using a zero-th order approximation, without the need to implement a different solver.
\end{remark}

There are two implementation choices that underlie the \code{solvers} module. First of all, solvers are designed to solve a static problem, since in \tvopt we model a time-varying problem as a sequence of static, sampled ones. We recall that dynamic costs provide the \code{sample} method. As a by-product, this also allows to employ \tvopt for prototyping and benchmarking static optimization algorithms.

\noindent The second design choice is to define solvers as functions, rather than objects, in order to provide a more flexible and efficient implementation. Indeed, in the course of solving~\eqref{eq:discrete-time-problem} a solver will be applied to several static problems and (possibly) using different parameters for each of them. As a consequence, defining a solver object is not very different from using a function, since the attributes (\emph{e.g.} problem and parameters) of the object would need to be changed often, cluttering the syntax.

\paragraph*{Metrics}
The sub-module \code{utils} provides the implementation of different metrics for evaluating the performance of a solver. Let $\{ \x_k \}_{k \in \N}$ be the sequence generated by a solver applied to~\eqref{eq:discrete-time-problem}. The available metrics are: \emph{fixed point residual} defined as $\{ \norm{\x_k - \x_{k-1}} \}_{k \geq 1}$; \emph{tracking error} defined as $\{ \norm{\x_k - \x^*(t_k)} \}$; and \emph{regret} defined as $\{ \frac{1}{k + 1} \sum_{j = 0}^{k} F(\x_j; t_j) - F(\x^*(t_j); t_j) \}_{k \in \N}$.

\paragraph*{Built-ins}
Examples of built-in solvers are gradient method, proximal point algorithm, forward-backward\footnote{Also called proximal gradient method.} and Peaceman-Rachford splittings, dual ascent, ADMM. The documentation \cite{documentation} lists all solvers with appropriate references.

\subsection{Constrained optimization}
We conclude this section by discussing how \tvopt can be used to solve online constrained optimization problems.

A first class of constraints that can be implemented is $\x \in \mathbb{C}$ where $\mathbb{C}$ is a non-empty, closed, and convex set. Indeed, given a \code{Set} object representing $\mathbb{C}$, we can define the indicator function of the set using an \code{Indicator} cost object. The indicator is $0$ for $\x \in \mathbb{C}$ and $+\infty$ for $\x \not\in \mathbb{C}$, and its proximal operator coincides with a projection onto $\mathbb{C}$.

\noindent Indicator functions can for example appear as the non-smooth term $g$ in a composite optimization problem $F(\x;t) = f(\x;t) + g(\x;t)$.

Equality and inequality constraints $g_i(\x;t) = 0$, $i = 1, \ldots, m$, and $h_i(\x;t) \leq 0$, $i = 1, \ldots, p$ can also be defined making use of \code{Cost} objects. The costs can then be used to define the Lagrangian of the constrained problem in order to apply primal-dual solvers \cite{boyd_convex_2004}.

A particular class of constrained problems that can be modeled using \tvopt is the following:
\begin{equation}\label{eq:admm-problem}
\begin{split}
	\x^*(t_k), \y^*(t_k) &= \argmin_{\x \in \R^n, \y \in \R^m} \left\{ f(\x; t_k) + g(\y; t_k) \right\} \\
	&\text{s.t.} \ \Am \x + \Bm \y = \cv
\end{split}
\end{equation}
where $\Am \in \R^{p \times n}$, $\Bm \in \R^{p \times m}$, $\cv \in \R^p$. As discussed in \cite{bastianello_primal_2020} and references therein, problem~\eqref{eq:admm-problem} can be solved by formulating its dual and applying suitable solvers to it. \tvopt provides the following \emph{dual} solvers: dual ascent, method of multipliers, ADMM, and dual forward-backward splitting. Moreover, a distributed version of dual ascent and ADMM are also implemented.

\noindent Notice that the constraints data $\Am$, $\Bm$, $\cv$ can be defined using NumPy's \code{ndarrays}, owing to the fact that the unknowns of an optimization problem are modeled as compatible arrays.

\section{Distributed Online Optimization}\label{sec:distributed}
This section describes the features of \tvopt that allow to model and solve distributed online problems. As in the case of centralized problems, we consider a sequence of samples from~\eqref{eq:distributed-continuous-time-problem}:
\begin{equation}\label{eq:distributed-discrete-time-problem}
\begin{split}
	\x^*(t_k) &= \argmin_{x_i \in \R^n} \sum_{i = 1}^{N} F_i(x_i; t_k) \\
	&\text{s.t.} \ x_i = x_j,\ \text{if $i$, $j$ connected}.
\end{split}
\end{equation}

We remark that the Python framework \texttt{DISROPT} \cite{farina_disropt_2019} is available for distributed optimization in static settings. Although there is some overlap with \tvopt's features described in this section, differently from \texttt{DISROPT} our goal is to model \emph{time-varying} problems over networks.

\subsection{Networks}
The sub-module \code{networks} defines the \code{Network} objects that model the connectivity pattern of a multi-agent system cooperating towards the solution of~\eqref{eq:distributed-discrete-time-problem}. The network is created from a given adjacency matrix.

A network implements the exchange of information between agents via the methods \code{send} and \code{receive}. In particular, \code{send} is called specifying a sender, receiver, and the packet to be exchanged. After calling \code{send}, the transmitted packet can be accessed using \code{receive}, which by default performs a destructive read of the message.

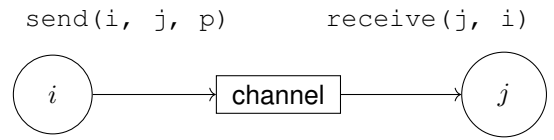
\begin{figure}[!ht]
\centering
\begin{tikzpicture}
		
	\node[agent] (i) at (0,0) {$i$};
	\node[agent] (j) at (6,0) {$j$};
	
	\node (ch) at (3,0) [draw,minimum width=1.65cm,minimum height=0.5cm] {\textsf{channel}};

	\path (i) edge[->] (ch) (ch) edge[->] (j);
	\node at (1,1) {\code{send(i, j, p)}};
	\node at (5,1) {\code{receive(j, i)}};
	
\end{tikzpicture}
\caption{A scheme representing agent-to-agent communication, with node \code{i} sending packet \code{p} to node \code{j}. The channel may be for example lossy or noisy.}
\end{figure}

The network implements also a \code{broadcast} method, using which an agents transmits the same message to all its neighbors. And a \code{consensus} method, which performs a consensus mixing of given local states using the \code{send} and \code{receive} methods.

In general, to define a new type of network (see \emph{e.g.} the built-ins) it is sufficient to overwrite the \code{send} method.

\paragraph*{Built-ins}
The built-in \code{Network} class models a standard loss-less network. Other types of networks available are a lossy network (transmissions may randomly fail), and noisy or quantized networks (which add Gaussian noise to or quantize the transmissions, respectively).

The sub-module provides also a number of built-in functions for generating the adjacency matrix of different types of graphs, for example: random, circulant or complete graphs.

\subsection{Costs}
Formulating distributed optimization problems is done using the \code{SeparableCost} object defined in \code{costs}, which models a cost $F : \R^{n_1 \times n_2 \times \ldots \times N} \times \R_+ \to \R \cup \{ +\infty \}$:
$$
	F(\x; t) = \sum_{i = 1}^{N} F_i(x_i; t)
$$
with $F_i$ the local cost function of the $i$-th agent. The cost is created from a list of static or dynamic local costs. Notice that the last dimension of $F$'s domain is the number of agents, using the flexibility of \code{Set} objects that allow for multiple dimensions.

\code{SeparableCost} implements all the methods described in section~\ref{subsec:costs}, with the difference that the outputs are arranged in an array with the last dimension indexing the agents. This choice allows for an easier access of the evaluation of each cost $F_i$. For example, if \code{F} is separable, then the result of \code{F.function(x, t)} will be $[F_1(x_1, t), \ldots, F_N(x_N, t)] \in \R^{1 \times N}$.

A \code{SeparableCost} also allows to evaluate \emph{e.g.} the gradient of a single component function $F_i$ by specifying the argument \code{i}.

\subsection{Solvers}
The sub-module \code{distributed_solvers} then provides built-in implementations of several distributed solvers. The difference with the centralized solvers described in section~\ref{subsec:solvers} is that these functions also require to be passed a \code{Network} object to perform agent-to-agent communications.

The built-in solvers are primal methods, \emph{e.g.} DPGM \cite{bastianello_distributed_2020}; primal-dual methods based on gradient tracking strategies, \emph{e.g.} \cite{shi_proximal_2015,alghunaim_decentralized_2020}; and dual methods, \emph{e.g.} dual decomposition and ADMM \cite{bastianello_asynchronous_2020}.

\noindent \tvopt also provides different functions to solve \emph{average consensus} using different protocols, for example gossip consensus \cite{aysal_broadcast_2009}.

\section{Numerical Examples}\label{sec:example}
The following sections presents a centralized example of online optimization benchmark and an example of distributed linear regression. A step-by-step discussion of the code is presented alongside some numerical results.

\subsection{Benchmark example}
The benchmark problem was proposed in \cite[section~IV.A]{simonetto_class_2016} and is characterized by
$$
	F(x;t) = (x - \cos(\omega t))^2 / 2 + \epsilon \log(1 + \exp(\varphi x))
$$
with $\omega = 0.02 \pi$, $\epsilon = 7.5$, and $\varphi = 1.75$. We test the prediction-correction framework (see \cite{bastianello_primal_2020}) using the extrapolation-based prediction $\hat{F}(\x;t_{k+1}) = 2 F(\x;t_k) - F(\x; t_{k-1})$\footnote{The alternative Taylor expansion-based prediction is also implemented in the examples section of \cite{code}.}.

\paragraph*{Setup}
Defining the cost requires fixing the sampling time $\Ts$ and a time horizon. 
\begin{python}
from tvopt import costs, prediction, solvers

# sampling time and time horizon
t_s, t_max = 0.1, 1e4

# cost function
f = costs.DynamicExample_1D(t_s, t_max)
\end{python}

\noindent We also define the simulation parameters, with \code{num_pred} and \code{num_corr} representing $\Np$ and $\Nc$. The solver we will use is a gradient method, so we define its step-size.

\begin{python}
# num. of prediction and correction steps
num_pred, num_corr = 5, 5

step = 0.2 # gradient method's step-size
\end{python}

\paragraph*{Prediction}
We define the extrapolation-based prediction (cf. Example~\ref{ex:prediction}) with
\begin{python}
f_hat = prediction.ExtrapolationPrediction(f, 2)
\end{python}
where the argument $2$ specifies the number of past costs to use for computing $\hat{f}$.

\paragraph*{Solution}
We then apply the prediction-correction solver as follows.
\begin{python}
x, x_hat = 0, 0

for k in range(f.time.num_samples):
    
    # correction
    p = {"f":f.sample(t_s*k)} # correction problem
    x = solvers.gradient(p, x_0=x_hat, step=step, 
                         num_iter=num_corr)
    
    # prediction
    f_hat.update(t_s*k)
    p = {"f":f_hat} # prediction problem
    x_hat = solvers.gradient(p, x_0=x, step=step, 
                             num_iter=num_pred)
\end{python}
In the code, we update $\x_k$ and $\hat{\x}_k$ during the correction and prediction steps, respectively. Moreover, notice that the dynamic cost $f$ is sampled every iteration, and that the prediction $\hat{f}$ is consequently updated. The correction and prediction problem are defined with a dictionary.

\smallskip

\begin{figure}[!ht]
\centering
\includegraphics[width=0.475\textwidth]{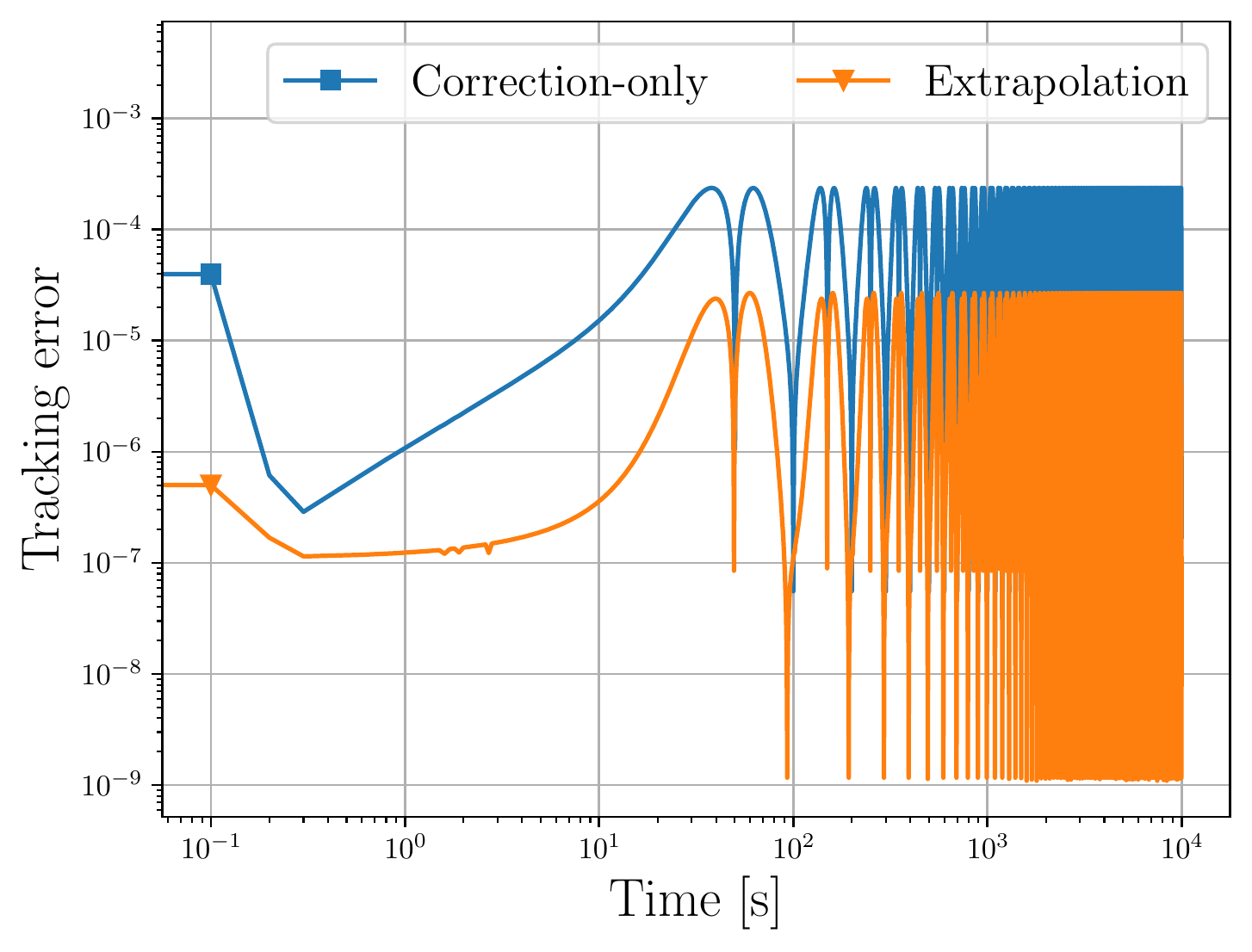}
\caption{Tracking error for different online optimization methods.}
\label{fig:tracking-error}
\end{figure}

Figure~\ref{fig:tracking-error} depicts the evolution of the tracking error $\{ \norm{\x_k - \x^*(t_k)} \}_{k \in \N}$ for the prediction-correction method discussed above. The method is compared with a correction-only strategy that does not employ a prediction to warm-start the solver at each sampling time.

The results show that prediction has a valuable effect in improving the performance of the online solver, and \tvopt provides an easy way to experiment with prediction strategies.

\subsection{Distributed linear regression}
The problem is~\eqref{eq:distributed-discrete-time-problem} with $N = 25$ agents and with local costs
$$
	F_i(x_i; t) = (a_i x_i - b_i(t_k))^2 / 2
$$
where $b_i(t_k) = a_i y(t_k) + e_i(t_k)$ and $y(t_k)$ is a sinusoidal signal with $e_i(t_k)$ a Gaussian noise of variance $\sigma^2 = 10^{-2}$. The solver employed is DGD \cite{yuan_convergence_2016}. We report a sample of the code in the following.

The network can be created as follows:
\begin{python}
# adjacency matrix
adj_mat = networks.random_graph(N, 0.5)
# network
net = networks.Network(adj_mat)
\end{python}
and the distributed, online solver is implemented with:
\begin{python}
x = x0 # initial condition

for k in range(f.time.num_samples):

    # problem creation
    problem = {"f":f.sample(t_s*k), "network":net}
    
    # distributed solver
    x = distributed_solvers.dpgm
          (problem, step, x_0=x, num_iter=num_iter)
\end{python}

\begin{figure}[!ht]
\centering
\includegraphics[width=0.475\textwidth]{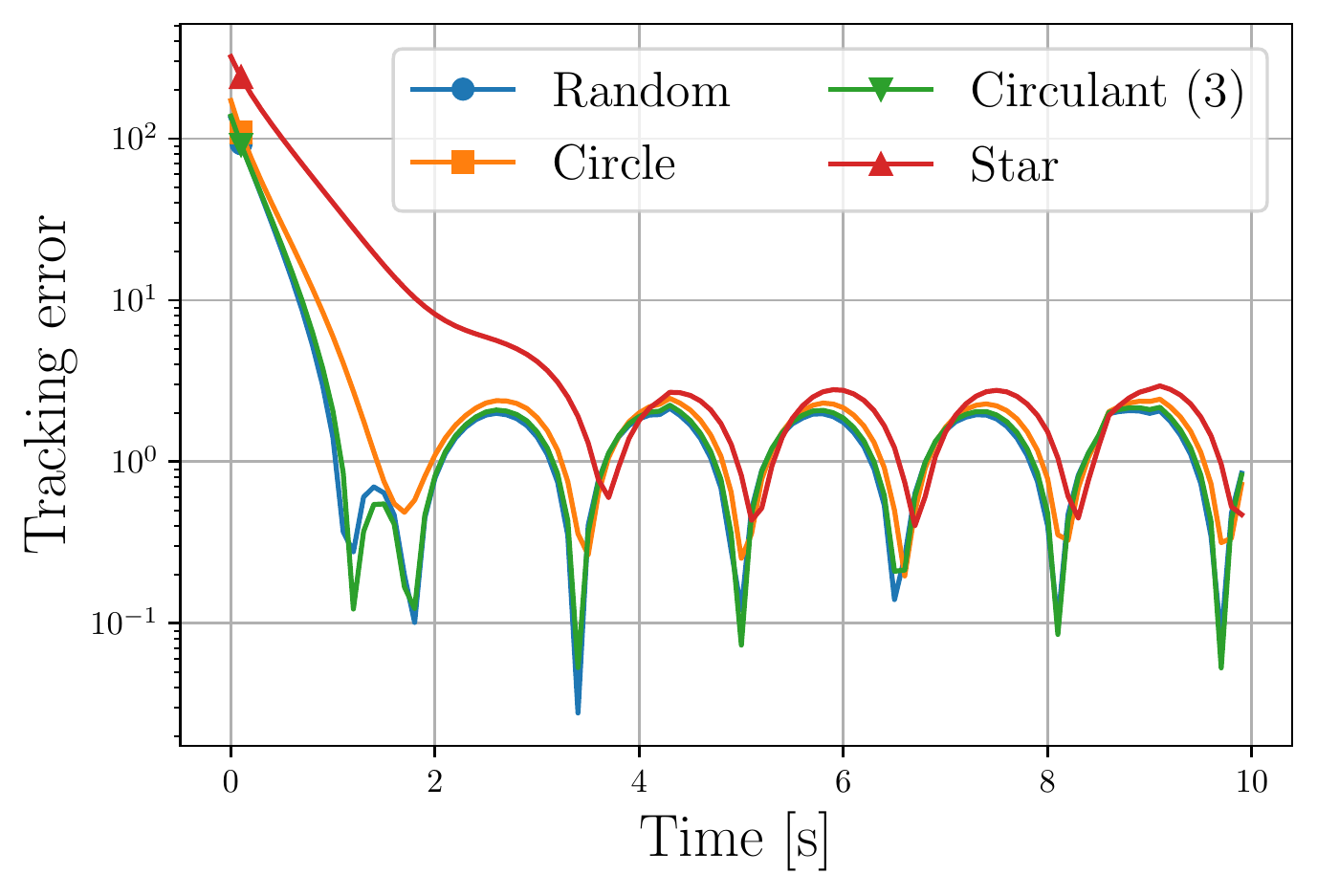}
\caption{Fixed point residual $\{ \norm{\x_k - \x_{k-1}}\}_{k \in \N}$ for different graph topologies.}
\label{fig:distributed}
\end{figure}

In Figure~\ref{fig:distributed} we report the fixed point residual (defined as $\{ \norm{\x_k - \x_{k-1}}\}_{k \in \N}$) for different graph topologies. We remark that the random graph has $\sim 225$ edges and thus is the more connected of the four topologies, which explains the fact that it achieves the better results.

\section*{Acknowledgment}
The author would like to thank Dr. Andrea Simonetto, Prof. Ruggero Carli, and Prof. Emiliano Dall'Anese for the many valuable discussions.

\bibliographystyle{IEEEtran}
\bibliography{IEEEabrv,references}


\end{document}